\documentclass[aps,prl,twocolumn,showpacs,preprintnumbers,amsmath,amssymb]{revtex4-1}
 
 \def\m{{\mu}}

 \def\frac#1#2{{#1\over #2}}

\def\be{\begin{equation}}
\def\ee{\end{equation}}
\def\ba{\begin{eqnarray}}
\def\ea{\end{eqnarray}}

\def\({\left(}
\def\){\right)}

\usepackage{color}
\usepackage{graphicx}
\usepackage{dcolumn}
\usepackage{bm}
\usepackage{ulem}

\begin{document}

\title{Probing QCD critical point and induced gravitational wave by black hole physics}

\author{Rong-Gen Cai$^{b,c}$, Song He$^{a,d}$, Li Li$^{b,c,e}$ and Yuan-Xu Wang$^a$}

\affiliation{$^a$Center for Theoretical Physics and College of Physics, Jilin University,\\ Changchun 130012, People's Republic of China}
\affiliation{$^{b}$CAS Key Laboratory of Theoretical Physics, Institute of Theoretical Physics, Chinese Academy of Sciences, Beijing 100190, China}
\affiliation{$^{c}$School of Fundamental Physics and Mathematical Sciences, Hangzhou Institute for Advanced Study, UCAS, Hangzhou 310024, China}
\affiliation{$^{d}$Max Planck Institute for Gravitational Physics (Albert Einstein Institute),\\ Am M\"uhlenberg 1, 14476 Golm, Germany}
\affiliation{$^e$Peng Huanwu Collaborative Center for Research and Education, Beihang University, Beijing 100191, China.}

\email{cairg@itp.ac.cn, \\hesong@jlu.edu.cn (corresponding author),\\liliphy@itp.ac.cn (corresponding author),\\yuanxu20@mails.jlu.edu.cn}

\date{\today}

\begin{abstract}
Locating the critical endpoint of QCD and the region of a first-order phase transition at finite baryon chemical potential is an active research area for QCD matter. We provide a gravitational dual description of QCD matter at finite baryon chemical potential $\mu_B$ and finite temperature using the non-perturbative approach from gauge/gravity duality. After fixing all model parameters using state-of-the-art lattice QCD data at zero chemical potential, the predicted equations of state and QCD trace anomaly relation are in quantitative agreement with the latest lattice results. We then give the exact location of the critical endpoint as well as the first-order transition line, which is within the coverage of many upcoming experimental measurements. Moreover, using the data from our model at finite $\mu_B$, we calculate the spectrum of the stochastic gravitational wave background associated with the first-order QCD transition in the early universe, which could be observable via pulsar timing in the future.

\end{abstract}

\maketitle

{\bf Introduction}--
As one of the most interesting and fundamental challenges of high energy physics, the phase diagram of QCD has been intensively studied. It involves the behavior of strongly interacting matter under extreme conditions~\cite{Braun-Munzinger:2008szb,Philipsen:2012nu,Gupta:2011wh}, ranging from cosmology and astrophysics to heavy-iron collisions. Due to the strong interaction in the non-perturbative regime, it has not been possible to obtain the full QCD phase diagram directly from QCD in terms of the temperature $T$ and the baryon chemical potential $\mu_B$ or the baryon number density $n_B$. While lattice QCD can provide reliable first-principles information at zero density~\cite{Borsanyi:2010cj,Borsanyi:2013bia,HotQCD:2014kol}, it fails at finite density due to the famous sign problem~\cite{Philipsen:2012nu}. Note, however, that the lattice data can be extrapolated to finite $\mu_B$ via different systematical schemes~\cite{Allton:2002zi,JETSCAPE:2020shq,Borsanyi:2021sxv}, but this is only reliable for small $\mu_B$.
Meanwhile, many effective low energy models have been proposed to study the QCD phase diagram under certain conditions, such as~\cite{Braun-Munzinger:2015hba,Stephanov:2004wx,Fukushima:2013rx}, for which it is difficult to match the lattice QCD data quantitatively. Nevertheless, the QCD matter is now believed to be in a hadronic phase of color-neutral bound states at low $T$ and small $\mu_B$, while it is deconfined at high $T$ and large $\mu_B$, known as the quark-gluon plasma. These two phases are separated by a transition at small $\mu_B$ and are expected to change into a first-order transition for higher $\mu_B$. The critical point between them is the QCD critical endpoint (CEP), which has been under active investigation by experiments and theoretical calculations.

On the other hand, the gauge/gravity duality~\cite{Maldacena:1997re,Gubser:1998bc,Witten:1998qj,Witten:1998zw} provides a powerful non-perturbative approach to solving the strongly coupled non-Abelian gauge theories by mapping to a weakly coupled gravitational system with one higher dimension. In particular, it provides a convenient way to incorporate real-time dynamics and transport properties at finite temperatures and densities. Previous studies (\emph{e.g.}~\cite{Gubser:2008yx,Gursoy:2008bu,Panero:2009tv,Gubser:2008yx,Jarvinen:2021jbd,Rougemont:2015ona}) have provided a strong indication that holography can make quantitative description of the properties of QCD in the non-perturbative regime.
The construction of the QCD phase diagram in such a holographic approach was initiated by~\cite{DeWolfe:2010he,DeWolfe:2011ts}, where the Einstein-Maxwell-dilaton (EMD) theory was used to mimic properties in the QCD phase diagram. Since then, several attempts have been made toward the direction of the phase diagram in the $T-\mu_B$ plane (see \emph{e.g.}~\cite{Cai:2012xh,He:2013qq,Alho:2013hsa,Critelli:2017oub,Knaute:2017opk,Gursoy:2017wzz,Yang:2020hun,Grefa:2021qvt,Demircik:2021zll}). In the spirit of effective field theory, the model parameters of the bulk gravitational theory should be fixed by matching with lattice QCD results. Therefore, it is crucial to use up-to-date lattice simulation for reliable prediction at finite $\mu_B$.

In this Letter, we construct a holographic QCD (hQCD) model in which all parameters are fixed using state-of-the-art lattice QCD data~\cite{HotQCD:2014kol} at $\mu_{B}=0$, generated by highly improved stagger fermion action. Moreover, all thermodynamic quantities are computed directly from the holographic renormalization and the so-called thermodynamic consistency relations~\cite{footnote}. Our prediction for the thermodynamic observables at finite $\mu_B$ is in quantitative agreement with the latest lattice results that are available for $\mu_B/T < 3.5$~\cite{Borsanyi:2021sxv}. We also calculate both the chiral and gluon condensates. Remarkably, we show for the first time in holography that the gluon condensate agrees with the lattice simulation regarding the QCD conformal anomaly. We then manage to make precise predictions for the QCD phase diagram at finite $\mu_B$, in particular a first-order line and a CEP located at $T_C=105 \text{MeV}$ and $\mu_C=555 \text{MeV}$.
Interestingly, the location of our CEP can be checked in the near future by many important upcoming facilities. Moreover, the strong first-order phase transition (SFOPT) in the early universe is an important source of the stochastic gravitational wave (GW) background (see, \emph{e.g.}~\cite{Cai:2017cbj, Caprini:2019egz, Hindmarsh:2020hop, Bian:2021ini} and references therein). While various scenarios of new physics beyond the Standard Model of particle physics have been considered to engineer an SFOPT in the literature, our present model provides a scenario for phase transition GWs within the Standard Model.


\textbf{Holographic model}
--We begin with the five-dimensional EMD theory
\begin{eqnarray}\label{action1}
S=\frac{1}{2\kappa_N^2}\int d^{5}x \sqrt{-g} \Big[\mathcal{R}-\frac{1}{2}\nabla_\mu \phi \nabla^\mu \phi\nonumber\\
-\frac{Z(\phi)}{4}F_{\mu\nu}F^{\mu\nu}-V(\phi)\Big]\,,
\end{eqnarray}
with a minimal set of fields for capturing the essential dynamics. Here $\kappa_{N}^{2}$ is the effective Newton constant. In addition to the metric $g_{\mu\nu}$ that characterizes the geometry, the real scalar $\phi$ (known as dilaton) encodes the running of the gauge coupling, and the Maxwell field $A_\mu$ accounts for a finite baryon density. $Z(\phi)$ and $V(\phi)$ are two phenomenological terms that will be fixed by matching to the lattice QCD at $\mu_B=0$.

The hairy black hole reads
\begin{equation}\label{phiansatz}
\begin{split}
ds^2=-f(r) e^{-\eta(r)} dt^2+\frac{dr^2}{f(r)}+r^2d\boldsymbol{x_3^2}\,,\\
\phi=\phi(r),\quad A_t=A_t(r)\,,
\end{split}
\end{equation}
where $d\boldsymbol{x_3^2}=dx^2+dy^2+dz^2$ and $r$ is the holographic radial coordinate with the asymptotical anti-de Sitter (AdS) boundary at $r\rightarrow\infty$.
Denoting the event horizon as $r_h$ at which $f$ vanishes, the temperature and entropy density can be obtained as
\begin{equation}
T=\frac{1}{4\pi}f'(r_h)e^{-\eta(r_h)/2},\quad s=\frac{2\pi}{\kappa_N^2} r_h^3\,,
\end{equation}
while the baryon chemical potential $\mu_B$ and density $n_B$ can be obtained from $A_t$ at the AdS boundary, we read the energy density $\epsilon$ and pressure $P$ directly by the dual stress-energy tensor via the holographic renormalization~\cite{Skenderis:2002wp,deHaro:2000vlm}. We refer to the Supplemental Material (SM) for more details~\cite{sm}. Then, the equation of state (EOS) and transport properties can be determined precisely. The form of $V$  and $Z$ is partially motivated by the one in~\cite{Gubser:2008yx,DeWolfe:2010he,DeWolfe:2011ts}, although these models were not able to simultaneously fit the lattice data for equilibrium and near-equilibrium features quantitatively.
\begin{figure}[hbt!]
	\centering
	\includegraphics[width=0.47\textwidth]{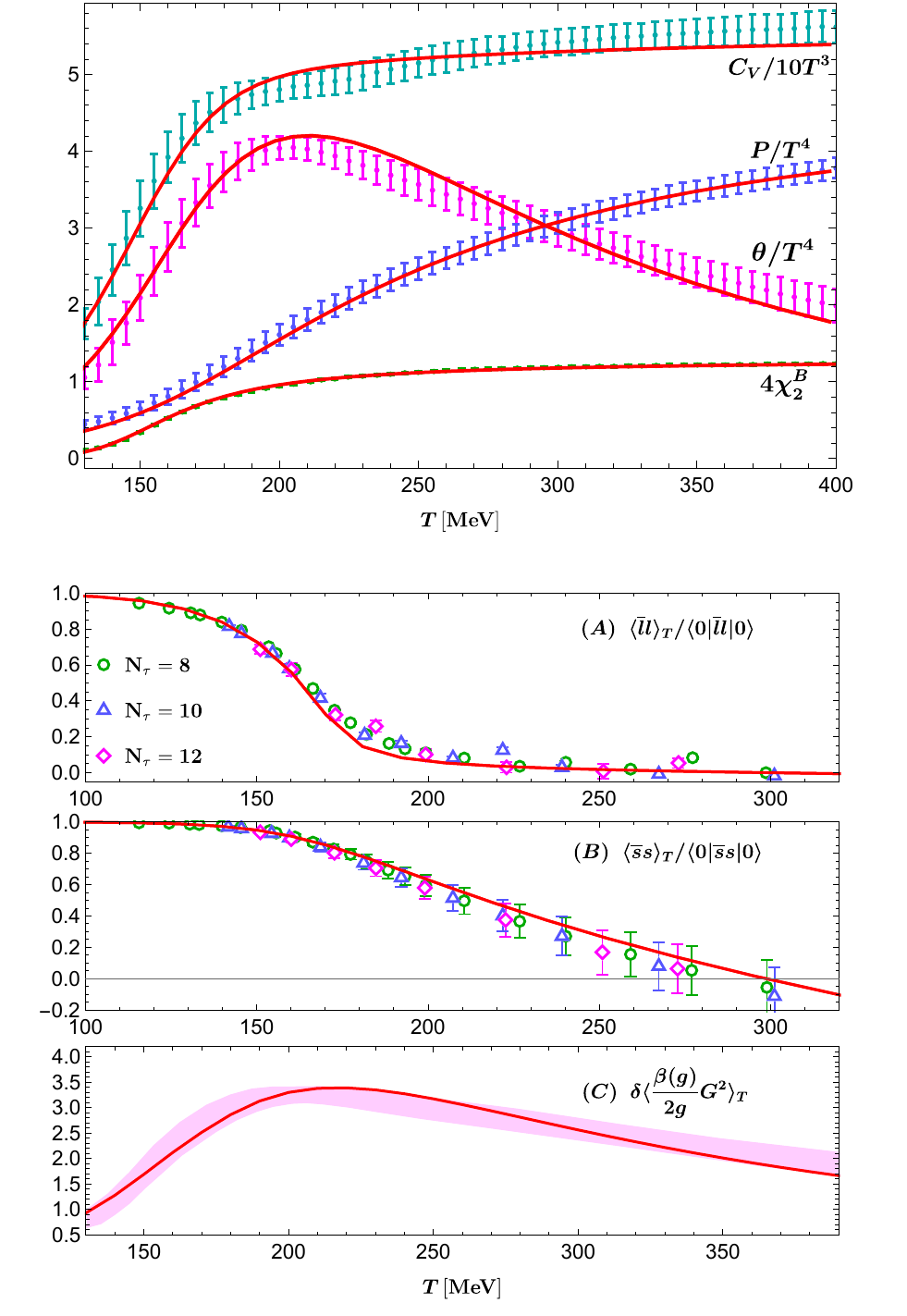}
	\caption{Thermodynamics and condensates at $\mu_B = 0$. \textbf{Upper panel:} The pressure $P$, the trace anomaly $\theta\equiv\epsilon -3P$, the specific heat $C_V$, and the baryon susceptibility $\chi_2^B$. \textbf{Lower panel:} The condensates for $(A)$ the $u, d$ quarks, $(B)$ the s quark, and $(C)$ the gluon. Data with error bars and the pink band are from lattice QCD~\cite{HotQCD:2014kol,Borsanyi:2021sxv}, while solid red curves are from our hQCD model. $N_\tau$ is the temporal extent in lattice QCD. {Following lattice QCD notations, the condensates of the light $(l=u,d)$ quarks and the $s$ quark are normalized by there values at zero temperature. The gluon condensate $\langle\frac{\beta(g)}{2g}G^2\rangle_T$ is subtracted by its vacuum value, where ${\beta(g)}$ is the $\beta$-function and $g$ is QCD gauge coupling.}}
\label{fig:lattice_comparison0}
\end{figure}
\begin{figure*}[hbt!]
	\centering
	\includegraphics[width=0.38\textwidth]{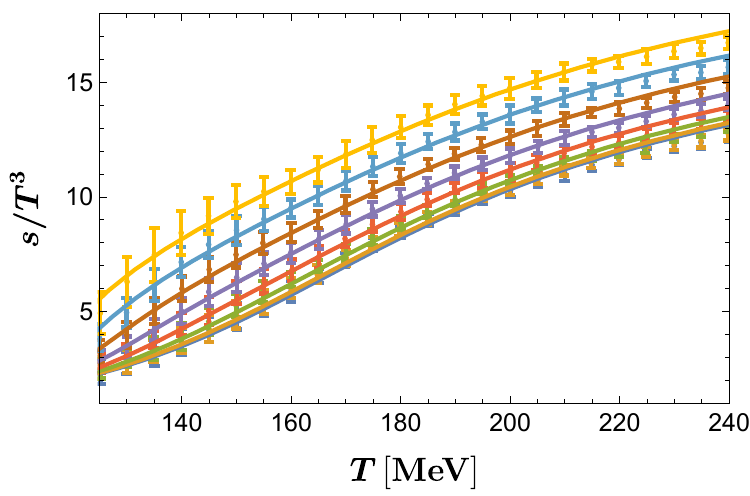} \quad
	\includegraphics[width=0.38\textwidth]{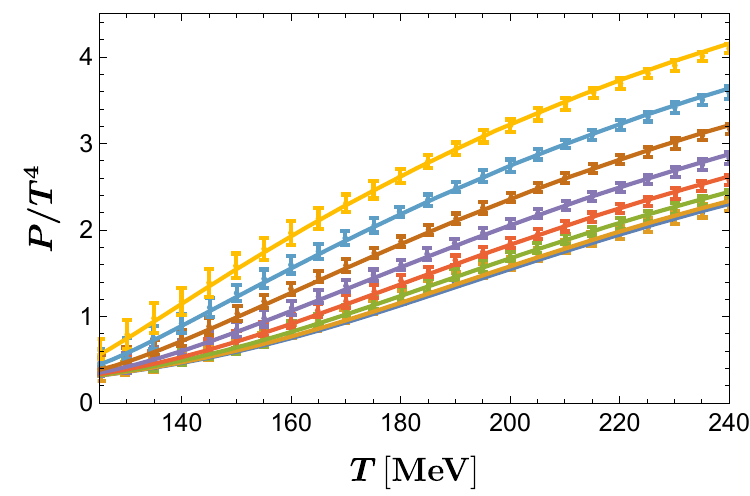}
	\includegraphics[scale=0.44]{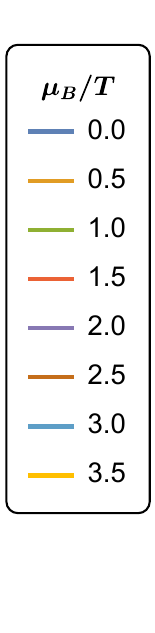}
	\caption{The entropy density $s$ (left) and the pressure $P$(right) at small chemical potentials. Our holographic computations (solid curves) are compared with the latest lattice QCD results from~\cite{Borsanyi:2021sxv}.}
	\label{fig:lattice_comparison1}
\end{figure*}

By global fitting the state-of-the-art lattice data~\cite{HotQCD:2014kol,Borsanyi:2021sxv} with (2+1)-flavors at zero net-baryon density (see Fig.~\ref{fig:lattice_comparison0}), the hQCD model can be fixed to be
\begin{equation}\begin{aligned}\label{vpossi}
V(\phi)&=-12\cosh[c_1\phi]+(6 c_1^2-\frac{3}{2})\phi^{2}+{c_2\phi^{6}}\,,\\
\kappa_{N}^{2} &=2 \pi({1.68}), \quad \phi_s={1085} \mathrm{MeV}\,, \\
Z(\phi) &=\frac{1}{1+c_{3}} {\text{sech}[c_4\phi^3]}+\frac{c_{3}}{1+c_{3}}e^{-c_{5} \phi}\,,
\end{aligned}\end{equation}
with $c_1=0.7100, c_2=0.0037, c_{3}=1.935, c_{4}=0.085, c_{5}=30$. Here $\phi_s=r \phi|_{r\rightarrow\infty}$ is the source term of the scalar field $\phi$, which essentially breaks the conformal symmetry and plays the role of the energy scale. The fitting results are presented in the upper panel of Fig.~\ref{fig:lattice_comparison0}. In addition to the EOS, the specific heat ${C_V}=(d\epsilon/d T)_{\mu_B}$, and the second-order baryon susceptibility $\chi_2^B=(dn_B/d\mu_B)_{T}/T^2$ at zero density, which are important quantities characterizing QCD transition, agree with the lattice data quantitatively.

Given the local bulk description of QCD~\eqref{vpossi}, we also calculate the gluon condensate and the condensates for $u, d, s$ quarks in the spirit of the KKSS model~\cite{Karch:2006pv} by introducing appropriate probe actions. The temperature dependence of these condensates is in quantitative agreement with the lattice simulations~\cite{HotQCD:2014kol}, see the lower panel of Fig.~\ref{fig:lattice_comparison0}. Remarkably, the trace anomaly $(\epsilon-3P)$ is found to be consistent with the summation of the quark and gluon condensates. It is the first time that this relation is realized in a holographic setup (see SM for technical details).

So far, our holographic model is completely fixed, and there are no free parameters. In Fig.~\ref{fig:lattice_comparison1}, we compare the holographic results with the latest lattice QCD~\cite{Borsanyi:2021sxv} which combines the Taylor-expanded approach and the analytical continuation approach. The holographic predictions are in quantitative agreement with the lattice results available for small chemical potentials, which strongly supports our hQCD model.

\begin{figure}[hbt!]
	\centering
	\includegraphics[width=0.47\textwidth]{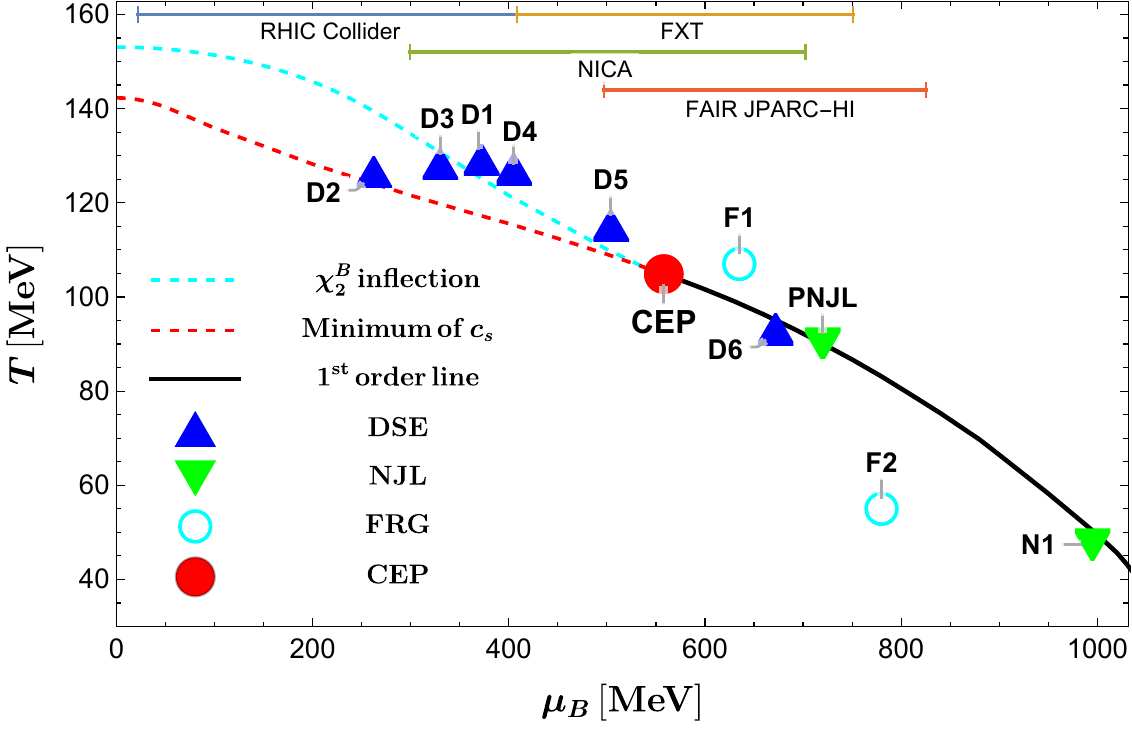}
	\caption{The QCD phase diagram predicted from our model. The minimum of {$c_s$} and the maximally increasing point of the $\chi_2^B$ are denoted by dashed red and cyan lines, respectively. The first-order phase transition line (solid black line) is determined by free energy. Our result for the location the CEP is $T_C=105 \text{MeV}, \mu_C=555 \text{MeV}$ (bold red dot). The CEP obtained by different approaches are presented as well, including Schwinger–Dyson equation (DSE), Nambu–Jona-Lasinio model (NJL), and Functional renormalization group (FRG). DSE: D1-D6 are from~\cite{Xin:2014ela,Gao:2016qkh,Qin:2010nq,Shi:2014zpa,Fischer:2014ata,Gao:2020qsj}. NJL: PNJL is from~\cite{Li:2018ygx} and NJL1 from~\cite{Asakawa:1989bq}. FRG: F1-F2 are from~\cite{Fu:2019hdw,Zhang:2017icm}. The coverage of RHIC~\cite{STAR:2020tga}, the STAR fixed target program (FXT), and future (FAIR, JPARC-HI, and NICA) experimental facilities~\cite{Fukushima:2020yzx} is also indicated at the top of the figure. 
	}
	\label{fig:phase diagram}
\end{figure}

\textbf{QCD phase diagram}--Having completely fixed the holographic model, we can now construct the phase diagram in the $T-\mu_B$ plane by numerically solving a series of black holes. We find that, as $\mu_B$ increases, the crossover on the $T$-axis is sharpened into a first-order line at the CEP (see Fig.~\ref{fig:phase diagram}). Since the transition from $\mu_B=0$ up to the CEP is a smooth crossover, there is no unique way to determine the transition temperature in the literature. Some suitable probes characterizing the drastic change of degrees of freedom between the quark-gluon plasma (QGP) and the hadron resonances gas are the minimum of the speed of sound $c_s$ and the maximum increasing point of $\chi_2^B$. The transition lines of the two probes are shown in Fig.~\ref{fig:phase diagram}. {The transition temperatures of the probes at zero $\mu_{B}$ compare well with the predictions of lattice QCD for the same up to the quark-hadron transition region around $140-160 \text{MeV}$~\cite{Bazavov:2009zn,Borsanyi:2010bp,Borsanyi:2011sw}}. Although they do not coincide quantitatively in the crossover region, they show similar behavior and are in the same order of magnitude. Moreover, they come together at the critical point $\mu_C$.

For a baryon chemical potential above $\mu_C$, the QCD transition becomes a first-order transition for which the free energy can uniquely determine the transition point from the holographic calculation (see the solid black line in Fig.~\ref{fig:phase diagram}). The first order line decreases monotonically with $\mu_B$. Although we cannot see when the line will terminate, our numerics suggest that the transition temperature becomes significantly small, approximately larger than {$\m_B\sim 1050 \text{MeV}$, beyond} which no stable black holes have been found. In this region, color superconductivity or related phenomena might be set in.

The CEP predicted by our hQCD model is located at $T_C=105 \text{MeV}, \mu_C=555 \text{MeV}$ (the red dot in Fig.~\ref{fig:phase diagram}). Therefore, there is no first-order phase transition in the region with $\mu_B/T<\mu_C/T_C\approx5.3$, which is consistent with the thermodynamics of Fig.~\ref{fig:lattice_comparison1}. We also show the location of CEP from different models and note that the current best lattice estimate suggests that CEP is likely above $\mu_B\sim 300\text{MeV}$~\cite{Stephanov:1999zu,Bazavov:2017dus}. The variation in priorities is considerable, but most of them are close to the transition line we found. Moreover, compared to other predictions, our CEP is close to those obtained by the Schwinger-Dyson equation~\cite{Fischer:2014ata,Gao:2020qsj} or the functional renormalization group~\cite{Fu:2019hdw} respectively. Nevertheless, all our predictions are at least qualitatively consistent with the consensus expectations for the QCD phase diagram. Interestingly, the predicted location of CEP is within the coverage of the STAR fixed target program (FXT) and future (FAIR ~\cite{CBM:2016kpk, Durante:2019hzd}, JPARC-HI and NICA~\cite{Sissakian:2009zza}) experimental facilities~\cite{Fukushima:2020yzx}. Therefore, our prediction can be verified in the near future.

\textbf{GWs from QCD phase transition}--An important feature associated with the first order QCD transition in the early universe is the generation of a stochastic background of GWs, for which bubbles of the broken phase nucleate and expand in the presence of a plasma background of false vacuum. The GWs can be generated in three processes: bubble collisions, sound waves, and magnetohydrodynamic turbulence. Some preliminary studies of the GW spectra from holographic QCD have been summarized in~\cite{Li:2021qer}. It has been recognized that higher-order corrections prevent true runaway transitions from occurring~\cite{Bodeker:2017cim}, thus the bubble wall terminal velocity $v_w<1$. For these non-runway bubbles, the GWs will be dominated by sound waves for which the energy spectrum reads~\cite{Guo:2021qcq}
\begin{equation}
\begin{split}
h^2\Omega_{GW}(f)=8.5\times 10^{-6}\left(\frac{100}{g_n}\right)^{1/3}\left(\frac{H_n}{\beta}\right)\left(\frac{\kappa \alpha}{1+\alpha}\right)^2\\
\times v_w \left(\frac{f}{f_{SW}}\right)^3\left[\frac{7}{4+3(f/f_{SW})^2}\right]^{7/2}\,,
\end{split}
\end{equation}
with the peak frequency
\begin{equation}
f_{SW}=1.9\times 10^{-8}\left(\frac{1}{v_w}\right)\left(\frac{\beta}{H_n}\right)\left(\frac{T_n}{100 \,\text{Mev}}\right)\left(\frac{g_n}{100}\right)^{1/6}\text{Hz}\,.
\end{equation}
The strength of the GW spectrum depends on many parameters that can be obtained from our holographic model. We discuss each of these quantities in turn. 

$T_n$ is approximately by the critical temperature of the first-order transition, see the solid black line in Fig.~\ref{fig:phase diagram}. Beyond the bag model approximation, the phase transition strength $\alpha$ is given by~\cite{Caprini:2019egz}
\begin{equation}\label{alpha}
\alpha=\frac{\theta_+-\theta_-}{3w_+}\Big{|}_{T=T_n}=\frac{\epsilon_+(T_n)-\epsilon_-(T_n)}{3w_+(T_n)}\,,
\end{equation}
between the false ($+$) and true ($-$) vacuums, where $\theta=\epsilon-3P$ is the trace anomaly, and $w=\epsilon+P$ is the enthalpy. Note that the numerator of~\eqref{alpha} is the latent heat for the first-order QCD phase transition at $T_n$. The effective number of degrees of freedom $g_n=45 s_+/(2\pi^2 T_n^3)$. We fix $v_w$ to the typical values $v_w=0.95$ for which the kinetic energy efficiency coefficient $\kappa=\frac{\alpha}{0.73+0.083\sqrt{\alpha}+\alpha}$. The only free parameter is the inverse time duration of the phase transition $\beta/H_n$.

\begin{figure}[h!]
\begin{center}
\includegraphics[width=0.47\textwidth]{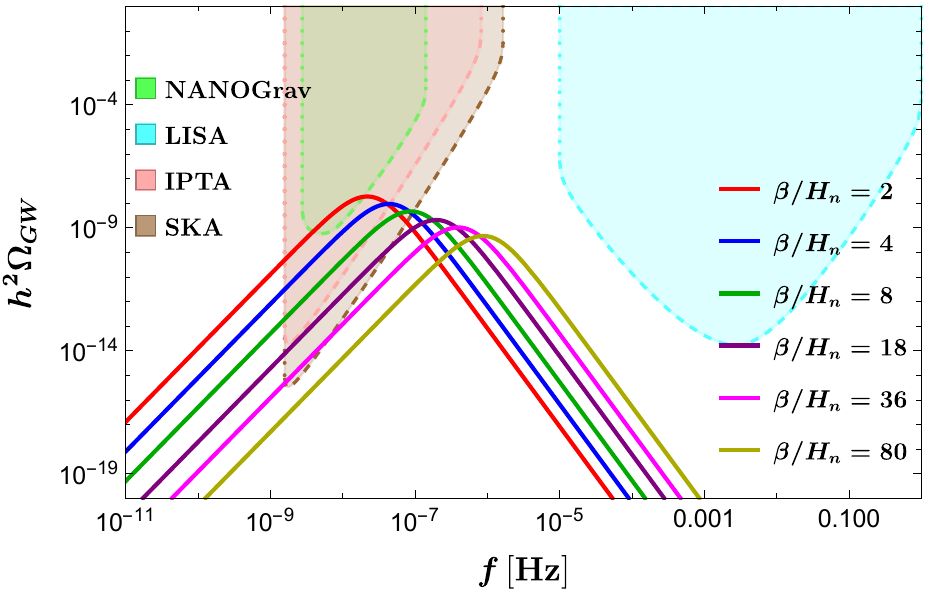}
\caption{Stochastic GW spectra predicted from the first-order QCD phase transition at $\mu_B=1000\,\text{MeV}$ where the baryon asymmetry matches the observation today. Parameters extracted from our holographic model are $(T_n,\alpha, g_n)=(49.53\,\text{MeV}, 0.33, 185)$ with the variation of $\beta/H_n$. The amplitude of GWs decreases as $\beta/H_n$ is increased. The sensitivities of four GW detectors (NANOGrav, LISA, IPTA and SKA)  are displayed~\cite{Schmitz:2020syl}.}
\label{fig:GW}
\end{center}
\end{figure}

Measurements of primordial element abundances and the anisotropy spectrum of the Cosmic Microwave Background (CMB) yield a strong constraint on the baryon sector, characterized by the baryon asymmetry $\eta_B\equiv n_B/s$. The observation value $\eta_B^{ob}\approx 10^{-10}$~\cite{Planck:2018vyg}. This quantity is conserved except during baryogenesis and a first-order phase transition. Thus, to apply to the cosmological QCD phase transition, the baryon asymmetry after the first-order transition in our model should be compatible with $\eta_B^{ob}$ in the later evolution of the universe. Near the CEP, our model yields $\eta_B\sim 10^{-2}$, which is eight orders of magnitude larger than the observation. Nevertheless, the value of $\eta_B$ for the true vacuum at the first-order phase transition line decreases significantly as $\mu_B$ increases. In particular, around $\mu_B=1000\text{Mev}$, $\eta_B$ reaches the same order of magnitude as the cosmological observation.

We therefore show the GW energy spectrum for $\mu_B=1000\text{Mev}$ ($\mu_B/T\approx 20$) in Fig.~\ref{fig:GW}, for which the cosmological QCD phase transition is first order and does not violate the constraint of a tiny baryon asymmetry. This transition point is very close to the CEP by NJL model~\cite{Asakawa:1989bq} (the green triangle at the bottom right of Fig.~\ref{fig:phase diagram}). To account for the uncertainty in the duration $\beta/H_n$, we will scan the space with $2\leq\beta/H_n\leq 80$. The energy density of GWs can reach $10^{-9}$ around $10^{-7}$Hz. While out of reach of the Laser Interferometer Space Antenna (LISA), it can be detected by the International Pulsar Timing Array (IPTA) and the Square Kilometer Array (SKA). The detection from the North American Nanohertz Observatory for Gravitational Wave (NANOGrav) is possible for extreme scenarios with small values of $\beta/H_n\sim\mathcal{O}(1)$.

Moreover, the QCD first-order phase transition with $\mu_B<1000\text{Mev}$ could also be realized in the early universe by considering the ``little inflation"~\cite{Boeckel:2009ej,Boeckel:2011yj} (see \emph{e.g.}~\cite{Linde:1985gh,kaempfer:1986,Borghini:2000} for earlier work and the SM). 
In this scenario, the large baryon asymmetry, which can be generalized by the well-established Affleck-Dine baryogenesis~\cite{Affleck:1984fy} is diluted by a short period of inflation in which the universe is trapped in a false vacuum state of QCD. We find that the e-folding number sufficient to dilute the baryon number in the little inflation scenario is at most $7$. Another scenario for a first-order QCD transition in agreement with observation is to consider a large lepton asymmetry~\cite{Schwarz:2009ii,Gao:2021nwz}, which we leave for future work.

\textbf{Discussion}--We have built a holographic model to confront the phase diagram in (2+1)-flavor QCD. The thermodynamics behaviors (EOS, transport, and condensates) are quantitatively matched with the latest lattice QCD simulation. The model has captured the characteristic QCD transition properties and offered a reliable first-order transition line and CEP in the QCD phase diagram at finite $\mu_B$. The predicted location of CEP and the GW energy spectrum of SFOPT from our hQCD model is within the detectivity of upcoming experimental facilities and, therefore, could be verified in the future. Moreover, our EOS at finite density could be crucial for supporting the experimental programs towards QCD matter, \emph{e.g.} heavy-ion collisions.

Dedicating to a precise characterization of QCD matter, particularly the properties and differences of the quark-gluon plasma and hadronic phases along the first-order phase transition, is an interesting direction for further study. We have set up the preliminary hQCD model to investigate the phase transition in the QCD phase diagram quantitatively. Many other relevant physical quantities should be considered to complete the phase diagram. Moreover, the present study should be embedded in the framework of a more general and multidimensional view of the QCD phase diagram, including magnetic field, isospin, and rotation. It will be interesting to consider real-time dynamics far from equilibrium.\\

\textbf{Acknowledgements}--We thank Heng-Tong Ding, Mei Huang, De-Fu Hou, Yi-Bo Yang, Danning Li, Zhibin Li, Shinya Matsuzaki, Shao-Jiang Wang, Yi Yang, Shou-Long Li and Peihung Yuan for stimulating discussions. This work is supported in part by the National Key Research and Development Program of China Grants No.2020YFC2201501 and No.2020YFC2201502, in part by the National Natural Science Foundation of China Grants No.12075101, No.12047569, N0.12122513, No.12075298, No.11991052, No.12235016 and No.12047503, in part by the Key Research Program of the Chinese Academy of Sciences (CAS) Grant NO. XDPB15 and by CAS Project for Young Scientists in Basic Research YSBR-006. S.H. would also like to appreciate the financial support from Jilin University and Max Planck Partner Group.

\section{Appendices}
\textbf{Equations of motion}--
From the action $(1)$ in the main text we can derive the equations of motion that are given as follows. 
\begin{equation}\label{eomps}
\begin{split}
\nabla_\mu \nabla^\mu\phi-\frac{\partial_\phi Z}{4} F_{\mu\nu}F^{\mu\nu}-\partial_\phi V&=0\,,\\
\nabla^\nu(Z F_{\nu\mu})&=0\,,\\
\mathcal{R}_{\mu\nu} -\frac{1}{2}\mathcal{R}g_{\mu\nu}= \frac{1}{2}\partial_\mu\phi \partial_\nu\phi&+\frac{Z}{2} F_{\mu\rho}{F_\nu}^\rho
\\+\frac{1}{2}\big(-\frac{1}{2}\nabla_\mu \phi \nabla^\mu \phi -\frac{Z}{4}F_{\mu\nu}F^{\mu\nu}&-V\big)g_{\mu\nu}\,.
\end{split}
\end{equation}
As the dual system lives in a spatial plane, we choose the Poincar\'{e} coordinates with $r$ the radial direction in the bulk. The metric ansatz reads
\begin{equation}\label{appansatz}
ds^2=-f(r) e^{-\eta(r)} dt^2+\frac{dr^2}{f(r)}+r^2(dx^2+dy^2+dz^2)\,,
\end{equation}
together with
\begin{equation}
\phi=\phi(r),\quad A_t=A_t(r)\,.
\end{equation}
We denote the event horizon as $r_h$ at which $f$ vanishes. Then the temperature and entropy density are given by
\begin{equation}
T=\frac{1}{4\pi}f'(r_h)e^{-\eta(r_h)/2},\quad s=\frac{2\pi}{\kappa_N^2} r_h^3\,.
\end{equation}

Substituting the ansatz into~\eqref{eomps}, we obtain the following independent equations of motion. 
\begin{equation}\label{eoms}
\begin{split}
\phi''+\left(\frac{f'}{f}-\frac{\eta'}{2}+\frac{3}{r}\right)\phi'+\frac{\partial_{\phi}Z}{2f}e^{\eta}A_t'^2-\frac{1}{f}\partial_\phi V& = 0\,, \\
\partial_r (e^{\eta/2} \, r^3\,Z  A_t')&=0\,,\\
\frac{\eta'}{r}+\frac{1}{3}\phi'^2&=0\,,\\
\frac{2}{r}\frac{ f'}{f}-\frac{\eta'}{r}+\frac{Z}{3f}e^{\eta}A_t'^2+\frac{2}{3 f}V+\frac{4}{r^2}&=0\,,
\end{split}
\end{equation}
where the prime denotes the derivative with respect to $r$. Both $Z(\phi)$ and $V(\phi)$ will be determined by matching to the lattice QCD at $\mu_B=0$.

In what follows we will specify these two functions as
\begin{equation}\begin{aligned}\label{vpossi1}
V(\phi)&=-12\cosh[c_1\phi]+(6 c_1^2-\frac{3}{2})\phi^{2}+c_2\phi^{6}\,,\\
Z(\phi) &=\frac{1}{1+c_{3}} {\text{sech}[c_4\phi^3]}+\frac{c_{3}}{1+c_{3}}e^{-c_{5} \phi}\,,
\end{aligned}\end{equation}
where $c_1, c_2, c_3, c_4, c_5$ are free parameters of the model. These model parameters capture the physical properties of realistic QCD,\emph{e.g.} EOS and baryon susceptibility. Near the AdS boundary $r\rightarrow\infty$ where $\phi\rightarrow 0$, one has
\begin{equation}
\begin{split}
&Z(\phi)=1+\mathcal{O}(\phi)\,,\\
&V(\phi)=-12-\frac{3}{2}\phi^2+\mathcal{O}(\phi^4)\,.
\end{split}
\end{equation}
Therefore, the cosmological constant is given by $\Lambda=-6$ (the AdS radius $L=1$). 

To obtain the numerical solutions for $f(r), \eta(r), \phi(r)$ and $A_t(r)$, we should specify suitable boundary conditions at both the event horizon $r_h$ and the AdS boundary $r\rightarrow\infty$. The smoothness of the event horizon yields the following analytic expansion in terms of $(r-r_h)$ in the IR:
\begin{equation}\label{irexpand}
\begin{split}
f&=f_h(r-r_h)+\dots\,,\\ \eta&=\eta_h^0+\eta_h^1(r-r_h)+\dots\,,\\
A_t&=a_h(r-r_h)+\dots\,,\\ \phi&=\phi_h^0+\phi_h^1(r-r_h)+\dots\,.
\end{split}
\end{equation}
After substituting~\eqref{irexpand} into the equations of motion~\eqref{eoms}, one finds four independents coefficients $(r_h, a_h, \eta_h^0, \phi_h^0)$. On the other hand, near the AdS boundary, we obtain the following asymptotic expansion: 
\begin{equation}\label{uvexpand}
\begin{split}
\phi(r)& = \frac{\phi_s}{r}+\frac{\phi_v}{r^3}-\frac{\ln(r)}{6r^3}(1-6c_1^4)\phi_s^3+\mathcal{O}(\frac{\ln(r)}{r^5})\,. \\
\quad A_t(r)&=\mu_B-\frac{2\kappa_N^2n_B}{2r^2}-\frac{2\kappa_N^2n_B\,c_3\,c_5\,\phi_s}{3(1+c_{3})\,r^3}\\
&+\frac{2\kappa_N^2n_B\,\phi_s^2\,((1+c_3)^2-6(-1+c_3)\,c_3\,c_5^2)}{48(1+c_{3})^2\,r^4}\\
&+\frac{2\kappa_N^2n_B\,c_3\,c_5\,(-10\,c_5^2\,(1+(-4+c_3)\,c_3))\,\phi_s^3}{300(1+c_{3})^3\,r^5}\\
&+\frac{2\kappa_N^2n_B\,c_3\,c_5\,\big((7-12\,c_1^4)\,\phi_s^3-{60\,\,\phi_v}\big)}{300(1+c_{3})\,r^5}\\
&-\frac{2\kappa_N^2n_B\,c_3\,c_5\,\phi_s^3(-1+6\,c_1^4)\ln(r)}{30(1+c_{3})\,r^5}+\mathcal{O}(\frac{\ln(r)}{r^6})\,.\\
\eta(r)&=0+\frac{\phi_s^2}{6r^2}+\frac{(1-6c_1^4)\phi_s^4+72\phi_s\phi_v}{144 r^4}\\
&-\frac{\ln(r)}{12 r^4}(1-6c_1^4)\phi_s^4+\mathcal{O}(\frac{\ln(r)^2}{r^6})\,.\\
f(r)&=r^2\Big[1+\frac{\phi_s^2}{6r^2}+\frac{f_v}{r^4}-\frac{\ln(r)}{12 r^4}(1-6c_1^4)\phi_s^4\\
&+\mathcal{O}\Big(\frac{\ln(r)^2}{r^6}\Big)\Big]\,.
\end{split}
\end{equation}
Note that we have taken the normalization of the time coordinate at the boundary such that $\eta(r\rightarrow\infty)=0$. $\phi_s$ is the source of the scalar operator of the boundary theory, which essentially breaks the conformal symmetry and plays the role of the energy scale.

Before proceeding, we point out that the equations of motion~\eqref{eoms} have two independent scaling symmetries:
\begin{eqnarray}\label{scaling1}
t\rightarrow\lambda_t\,t\,,\,\,\,\,e^\eta\rightarrow\lambda_t^2\,e^\eta\,,\,\,\,\,A_t\rightarrow\lambda_t^{-1}A_t\,,\\\label{scaling2}
r\rightarrow\lambda_r\,r\,,\;\;\;\;\,\,f\rightarrow\lambda_r^2\,f\,\,,\,\,\,\,\,A_t\rightarrow\lambda_r\,A_t\,,
\end{eqnarray}
with $\lambda_t$ and $\lambda_r$ constants. Thus, we can first set $\eta_h^0=0$ and $r_h=1$ for performing numerics. After obtaining the numerical solutions, we should use the first symmetry to satisfy the asymptotic condition $\eta(r\rightarrow\infty)=0$ and use the second one to fix the energy scale $\phi_s$.\\

\textbf{Thermodynamics}--We now compute the free energy density $\Omega$ which is identified as the temperature $T$ times the renormalized action in the Euclidean signature.  Since we consider a stationary problem, the Euclidean action is related to the Minkowski one by a minus sign. Moreover, we should include the Gibbons-Hawking boundary term for a well-defined Dirichlet variational principle and a surface counterterm for removing divergence. Therefore, we have
\begin{equation}\label{freeE}
-\Omega V=T (S+S_\partial)_{on-shell}\,,
\end{equation}
with $V=\int dx dy dz$ and $t\in[0,1/T]$.
\begin{figure*}[hbt!]
	\centering
	\includegraphics[width=0.43\textwidth]{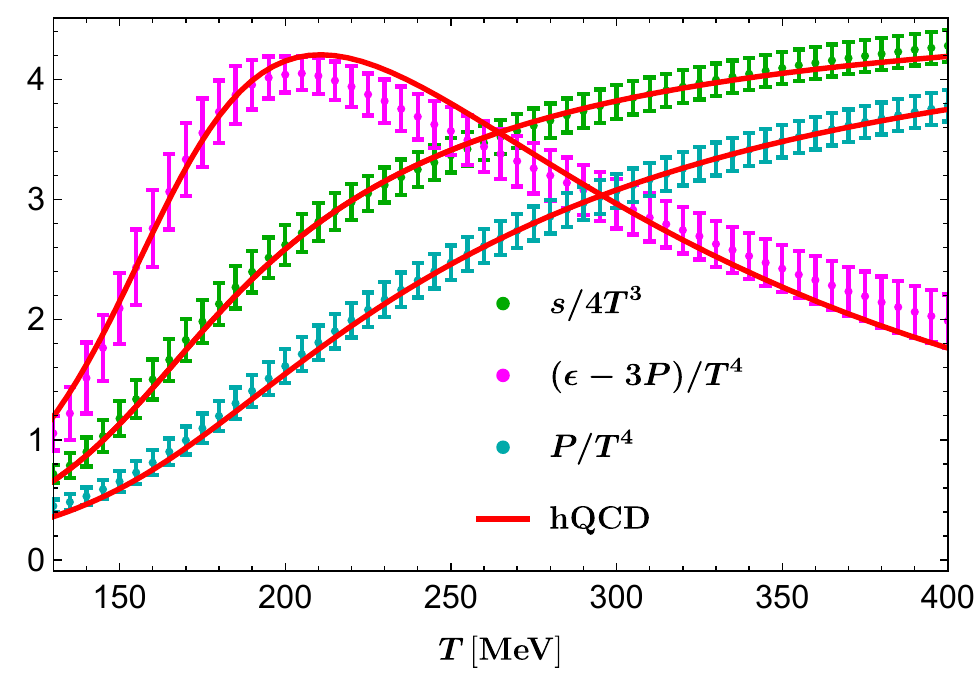}\qquad
	\includegraphics[width=0.43\textwidth]{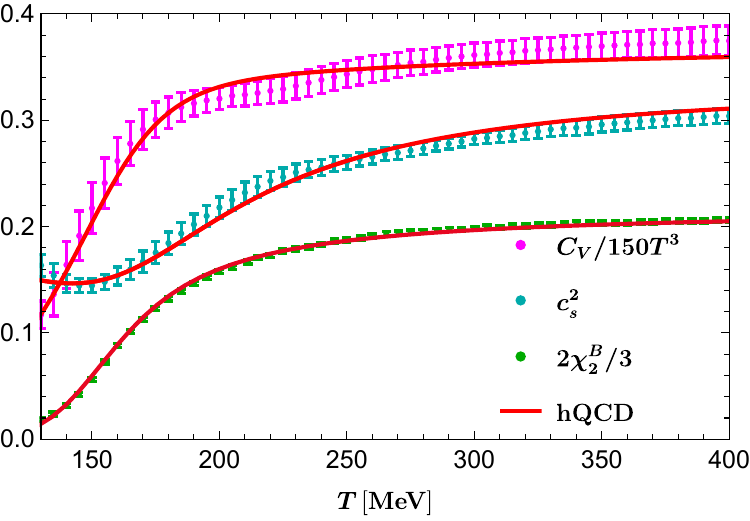}
	\caption{Thermodynamics at $\mu_B = 0$. Data with error bars are from the latest lattice QCD~\cite{HotQCD:2014kol,Borsanyi:2021sxv}, while solid curves from our hQCD model, including entropy density $s$, trace anomaly $(\epsilon -3P)$, pressure $P$, specific heat $C_V$, squared sound speed $c_s^2$ and baryon susceptibility.}
\label{fig:lattice_app}
\end{figure*}

We work in the grand canonical ensemble for which the baryon chemical potential is fixed. For the model~\eqref{vpossi1} we are considering, following the the holographic renormalization~\cite{Skenderis:2002wp,deHaro:2000vlm}, the boundary terms take the form 
\begin{eqnarray}\label{byterm}
S_\partial &=\frac{1}{2\kappa_N^2}\int_{r\rightarrow\infty}dx^4\sqrt{-h}\Big[2K-6-\frac{1}{2}\phi^2\nonumber\\&-\frac{6c_1^4-1}{12}\phi^4 \ln[r]-b\,\phi^4+\frac{1}{4}F_{\rho\lambda}F^{\rho\lambda}\ln[r]\Big]\,,
\end{eqnarray}
where $h_{\mu\nu}$ is the induced metric at the AdS boundary and $K_{\mu\nu}$ is the extrinsic curvature defined by the outward pointing normal vector to the boundary. 

Employing the equations of motion~\eqref{eoms} and the asymptotical expansion~\eqref{uvexpand}, we obtain
\begin{equation}\label{free}
\begin{split}
\Omega &=\lim_{r\rightarrow\infty}\Bigg[2 e^{-\eta/2}r^2f-e^{-\eta/2}r^3\sqrt{f}\Big(2K-6\\
&\qquad\qquad-\frac{1}{2}\phi^2-\frac{6c_1^4-1}{12}\phi^4\ln(r)-b\phi^4\Big)\Bigg]\,,\\
&=\frac{1}{2\kappa_N^2}\left(f_v-\phi_s\phi_v-\frac{3-48 b-8c_1^4}{48}\phi_s^4\right)\,.
\end{split}
\end{equation}
The energy-momentum tensor of the dual boundary theory reads
\begin{eqnarray}\label{Tmunu}
T_{\mu\nu}&=&\lim_{r\rightarrow\infty}\frac{2\,r^2}{\sqrt{-\operatorname{det} h}} \frac{\delta (S+S_\partial)_{on-shell}}{\delta h^{\mu \nu}},\nonumber\\
&=&\frac{1}{2\kappa_N^2}\lim_{r\rightarrow\infty}r^2\Big[2(K h_{\mu\nu}-K_{\mu\nu}-3 h_{\mu\nu})\nonumber\\&-&(\frac{1}{2}\phi^2+\frac{6c_1^4-1}{12}\phi^4 \ln[r]+b\,\phi^4)h_{\mu\nu}
\nonumber\\&-&(F_{\mu\rho}{F_{\nu}}^{\rho}-\frac{1}{4}h_{\mu\nu}F_{\rho\lambda}F^{\rho\lambda})\ln[r]\Big]\,,
\end{eqnarray}
Substituting~\eqref{uvexpand}, we obtain
\begin{equation}\label{EP}
\begin{split}
\epsilon &:= T_{tt}=\frac{1}{2\kappa_N^2}\left(-3 f_v+\phi_s\phi_v+\frac{1+48 b}{48}\phi_s^4\right)\,,\\
P&:=T_{xx}=T_{yy}=T_{zz}\\
&=\frac{1}{2\kappa_N^2}\left(-f_v+\phi_s\phi_v+\frac{3-48 b-8c_1^4}{48}\phi_s^4\right)\,,
\end{split}
\end{equation}
with vanishing non-diagonal components. One immediately finds that $P=-\Omega$, which is expected from thermodynamics. The trace of the energy-momentum tensor is given by
\begin{equation}
\epsilon-3P=\frac{1}{2\kappa_N^2}\left(2\phi_s\phi_v+\frac{1-24 b-3c_1^4}{6}\phi_s^4\right)\,.
\end{equation}
It is manifest that there is a trace anomaly in the presence of the source term $\phi_s$ that breaks the conformal symmetry.

By integrating the second equation of~\eqref{eoms}, one has
\begin{equation}
\frac{1}{2\kappa_N^2}e^{\eta/2} \, r^3\,Z  A_t'= n_B\,,
\end{equation}
where the constant $n_B$ is nothing but the charge density that can be computed using the standard holographic dictionary. Another useful radially conserved quantity reads~\cite{Kiritsis:2015hoa,Cai:2020wrp}
\begin{equation}
\mathcal{Q}=\frac{1}{2\kappa_N^2}r^3 e^{\eta/2}\left[r^2\left(\frac{f}{r^2}e^{-\eta}\right)'-Z A_t A_t' \right],
\end{equation}
which connects horizon to boundary data. Evaluating at the horizon, we obtain
\begin{equation}
\mathcal{Q}=T s,
\end{equation}
and therefore $\mathcal{Q}=0$ signals the extremity. On the other hand, evaluating $\mathcal{Q}$ at the AdS boundary, we find
\begin{equation}
\mathcal{Q}=\epsilon+P-\mu_B\,n_B\,.
\end{equation}
Therefore, we immediately obtain the expected thermodynamical relation 
\begin{equation}\label{free2}
\Omega=\epsilon-T\,s-\mu_B\,n_B=-P,
\end{equation}
where $\Omega$ is the free energy density~\eqref{free}.

\begin{figure*}[hbt!]
	\begin{center}
		\includegraphics[width=0.95\textwidth]{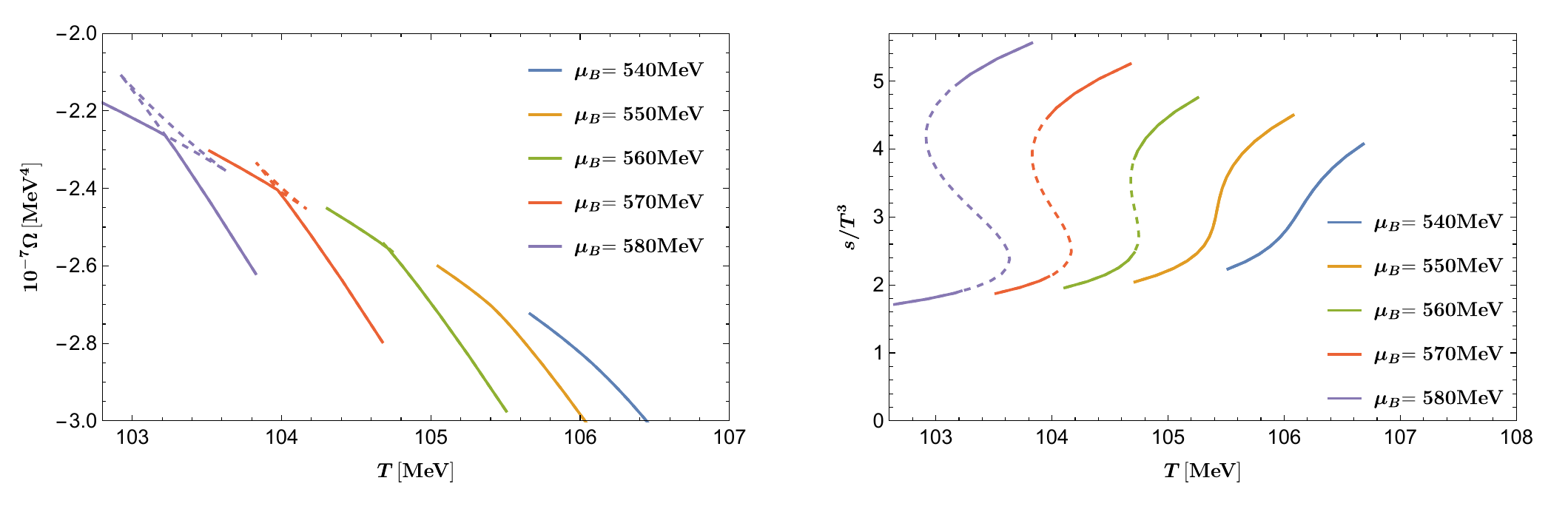}
		\caption{Illustration of the free energy $\Omega$ (left) and entropy density $s$ (right) for different values of $\mu_{B}$. From right to left, $\mu_{B}$ increases. For small $\mu_B$, there is no phase transition. For $\mu_B>\mu_{C}=555\text{MeV}$, both $\Omega$ and $s$ are multivalue functions of temperature, implying a first order phase transition. For each $\mu_B$, the thermodynamic favored states are denoted by solid lines, while the unfavored ones are dashed lines.}
	\label{fig:frees}
	\end{center}
\end{figure*}
\begin{figure}[hbt!]
	\centering
	\includegraphics[width=0.43\textwidth]{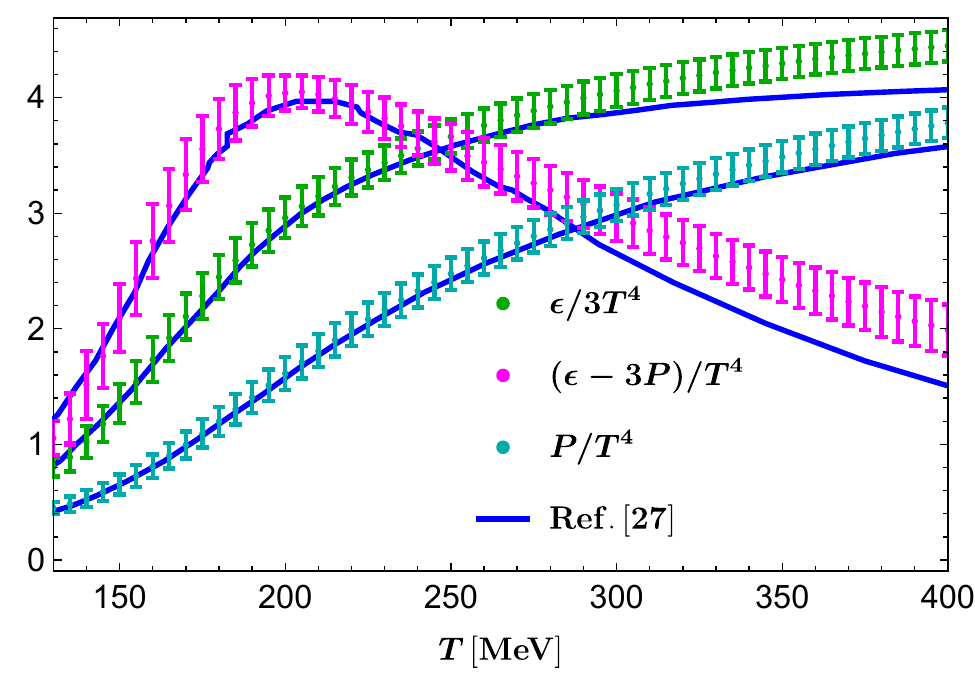}	
	\caption{Thermodynamics at $\mu_B = 0$. Data with error bars are from the latest Lattice QCD~\cite{HotQCD:2014kol}, while blue curves are from the holographic model of~\cite{Grefa:2021qvt}. The entropy density $\epsilon$, trace anomaly $(\epsilon -3P)$, and pressure $P$ fail to fit the lattice data for the temperature above $280\text{MeV}$.}
\label{fig:lattice_old}
\end{figure}

After obtaining the above thermodynamic quantities, we can also compute some important transport coefficients, including the sound speed $C_s=\sqrt{(dP/d\epsilon)_{\mu_B}}$, the specific heat $C_{V}=(d\epsilon/d T)_{\mu_B}$, and the second-order baryon susceptibility $\chi_2^B=(dn_B/d\mu_B)_{T}/T^2$. These properties are compared to the state-of-the-art lattice data~\cite{HotQCD:2014kol,Borsanyi:2021sxv} with (2 + 1)-flavors at zero baryon density, from which all free parameters of our hQCD model can be fixed, see Eq.~(4) and Figs.~1 and~2 in the main text. Moreover, the parameter $b$ that appears in the boundary terms~\eqref{byterm} is chosen to be $b=-0.27341$. We stress that $b$ is vital to build a hQCD model, which is necessary to satisfy the lattice QCD simulation for which $P(T=0)=0$ at zero baryon chemical potential. Without a concrete holographic renormalization, one cannot fix $b$. As far as we know, this important point has been overlooked in the literature. In most studies, the thermodynamic variables were obtained only by integrating the standard first law of black hole thermodynamics in a finite temperature range. We show the fitting results for more quantities in  Fig.~\ref{fig:lattice_app}. In addition to the EOS (left panel), the sound speed ${c_s}=\sqrt{(dP/d\epsilon)_{\mu_B}}$, the specific heat ${C_V}=(d\epsilon/d T)_{\mu_B}$, and the second-order baryon susceptibility $\chi_2^B=(dn_B/d\mu_B)_{T}/T^2$ at zero density, which are three important quantities characterizing QCD transition, agree with the lattice data quantitatively, see the right panel of Fig.~\ref{fig:lattice_app}.

In the real world, the quarks are massive, the chiral symmetry is not an exact symmetry of QCD, and the center symmetry is not exact. There are no well-established order parameters to probe the phase transition. It is non-trivial to identify whether the phase transition is associated with the chiral phase transition or deconfinement phase transition. Nevertheless, the defining feature for a first-order phase transition can be identified from the free energy computed directly in our holographic model, see~\eqref{free} and~\eqref{free2}. In Fig.~\ref{fig:frees}, we show the free energy density $\Omega$ and entropy density $s$ as a function of $T$ for different $\mu_B$. The temperature dependence of the free energy decreases smoothly for $\mu_B<\mu_C$. At the same time, it becomes a swallowtail for $\mu_B>\mu_C$, singling a first-order phase transition that has been widely used to mimic the phase transition in holographic QCD. The location of the CEP is found to be $\mu_c=555\text{MeV}$ and $T_c=105\text{MeV}$. The corresponding behavior of the entropy density versus $T$ is presented in the right panel of Fig.~\ref{fig:frees}. A first-order jump in the entropy density manifests when $\mu_B>\mu_C$.

\begin{figure*}[hbt!]
	\centering
	\includegraphics[width=0.43\textwidth]{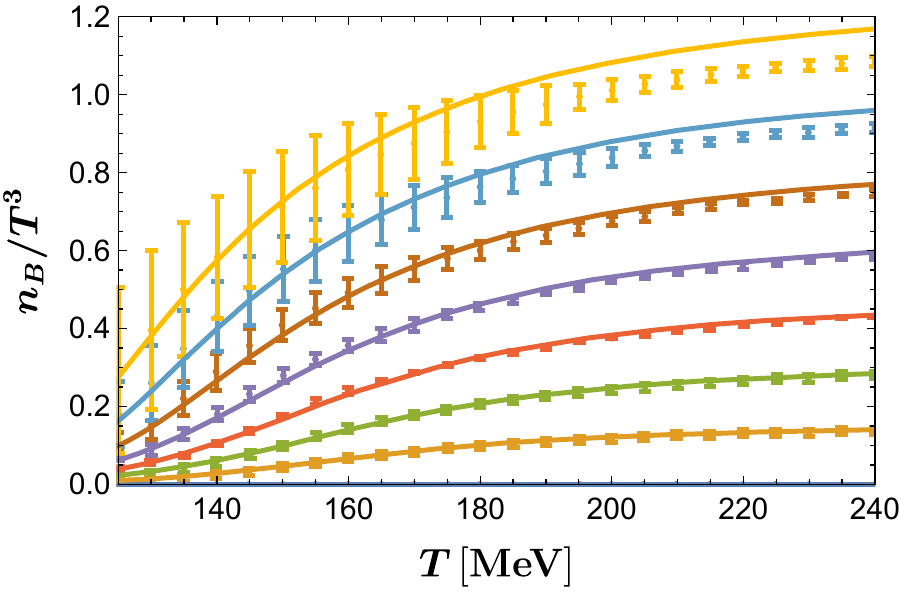}\quad
	\includegraphics[width=0.43\textwidth]{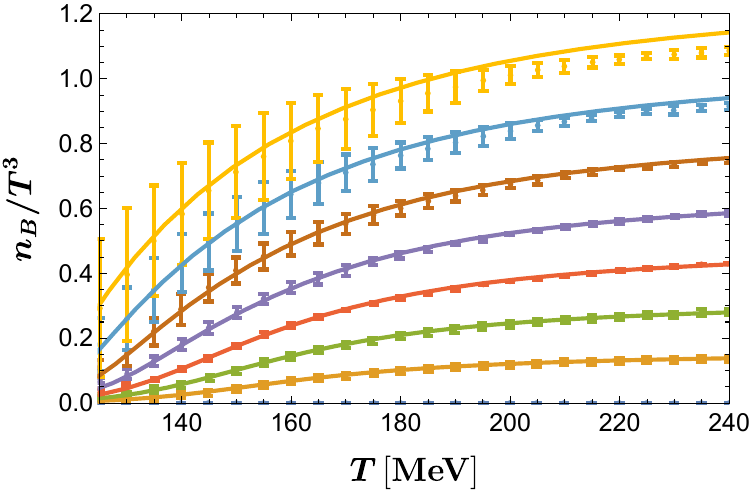}
	\includegraphics[scale=0.50]{legend.pdf}
	\caption{The baryon density $n_B$ at small chemical potentials. We compare the model of~\cite{Grefa:2021qvt} (left) and our model (right) with the latest lattice QCD results~\cite{Borsanyi:2021sxv}.}
	\label{fig:lattice_nB}
\end{figure*}
Before ending this section, we compare the thermodynamics of the holographic model~\cite{Grefa:2021qvt} to the state-of-the-art lattice QCD data~\cite{HotQCD:2014kol} used in the present work. This holographic model was constructed to mimic the QCD thermodynamics at a quantitative level. As can be seen from Fig.~\ref{fig:lattice_old}, the model of~\cite{Grefa:2021qvt} fails to fit the new lattice data for the temperature above $280\text{MeV}$ at $\mu_B=0$. Moreover, it is much more challenging to fit $n_B$. One can see from the left panel of Fig.~\ref{fig:lattice_nB} that the model of~\cite{Grefa:2021qvt} can only quantitatively match the latest QCD results for baryon density below $\mu_B/T= 2.0$. This fitting improves significantly in our model, see the right panel. Moreover, our setup is much simpler than the one in~\cite{Grefa:2021qvt}, which, as far as we know, was the only effective holographic model that agreed with the old lattice QCD data~\cite{Borsanyi:2013bia} for EOS and transport coefficients at the quantitative level.\\

\textbf{Quark and gluon condensates}--The non-perturbative quantum fluctuations generate quark and gluon condensates, which break chiral or dilatation symmetries of QCD. While there have been intensive studies in the literature, a complete and non-perturbative understanding of realistic QCD is still missing, and many features of these condensates are not yet well understood and established. Our holographic model has been shown to describe QCD thermodynamics quite well. A natural question is how to incorporate the flavor dynamics and gluon condensate in our framework. Following the spirit of the KKSS model~\cite{Karch:2006pv}, we introduce a holographic probe action to evaluate the chiral and gluon condensates. 

We consider the (2+1)-flavor QCD for which $m_u=m_d<m_s$ with $m_u$, $m_d$ and $m_s$ the mass for $u$, $d$ and $s$ quarks, respectively. Treating the flavor part as a probe, we introduce the effective form of the action

\begin{equation}\label{actionprobe}
S_{X_q}=\frac{1}{2\kappa_N^2}\int d^{5}x \sqrt{-g} \,Z_{q}(\phi)\big[-\frac{1}{2}\nabla_\mu X_q \nabla^\mu X_q-V(X_q)\big]\,,
\end{equation}
in the background~\eqref{appansatz},
where $q=l$ denotes the light quarks ($u$ and $d$) and $q=s$ is for the $s$ quark. The bulk field $X_{q}$ is dual to the boundary $\bar{\psi}_{q}\psi_{q}$ chiral operator with the scaling dimension $\Delta_c=3$. The effective coupling $Z_q(\phi)$ and the potential $V(X_q)$ will be fixed later by matching the lattice QCD data. Then, we need to solve the equation of motion derived from~\eqref{actionprobe}.
\begin{equation}\label{probeeom}
\begin{split}
X_q''+\left(\frac{f'}{f}-\frac{\eta'}{2}+\frac{3}{r}\right)X_q'+\frac{\partial_\phi Z_{q}\,X_q'\,\phi'}{Z_{q}}-\frac{1}{f}\partial_{X_q} V & = 0\,.
\end{split}
\end{equation}

Since the dynamics of $u, d$ and $s$ quarks in $2+1$ flavored QCD is quite different, we introduce $Z_l(\phi)$ and $Z_s(\phi)$ to distinguish the dynamics of light and heavy quarks. We choose $Z_q$ ($q=l, s$) and $V$ as follows
\begin{equation}
\begin{aligned}\label{vpossi11}
Z_{q}(\phi) &=a_{1}\,e^{a_{q}\,\,\phi^2}\,,\\
V(X_q) &=-\frac{3}{2}X_q^2+a_0\,X_q^4 \,,
\end{aligned}
\end{equation}
with $a_0, a_1, a_l, a_s$ constants. Note that the mass of the bulk field $X_q$ is determined by the scaling dimension of the dual operator $\bar{\psi}_{q}\psi_{q}$. Therefore, the UV expansion could be obtained as  
\begin{equation}\label{uvexpand1}
\begin{aligned}
X_{q}(r)&=\frac{m_{q}}{r}+\dots+\frac{\sigma_q}{r^3}+\dots\,,
\end{aligned}
\end{equation}
where $m_{l}=4.5\text{MeV}$ and $m_s=90\text{MeV}$ are the light and heavy quark masses, respectively~\cite{HotQCD:2014kol}. According to the holographic dictionary, the chiral condensates read
\begin{equation}
\begin{aligned}\label{vpossi12}
\langle\bar{\psi}\psi\rangle_{q,T}&=\frac{\delta S_{X_q}^{ren}}{\delta m_q}=\frac{a_1}{2 \kappa_N^2}\big[2\sigma_q+2a_0m_q^3+\frac{1}{4}m_q\phi_s^2\big]\,,
\end{aligned}
\end{equation}
where the renormalized action $S_{X_q}^{ren}\equiv S_{X_q}+S_{X_q,\partial}$ with the boundary counter term
\begin{equation}\begin{aligned}\label{vpossi13}
S_{X_q,\partial} &=\frac{1}{2\kappa_N^2}\int_{r\rightarrow\infty}dx^4\sqrt{-h}\Big[-\frac{1}{2}a_1X_q^2+a_1a_0X_q^4\ln{r}\\
&+\frac{a_1(1-6a_q)}{6}X_q^2\phi^2\ln{r}\Big]\,.
\end{aligned}
\end{equation}
By requiring the regularity of the solution at the event horizon, one can obtain the chiral condensates for light and heavy quarks in the function of $T$.

\begin{figure*}[hbt!]
\begin{center}
\includegraphics[width=1.0\textwidth]{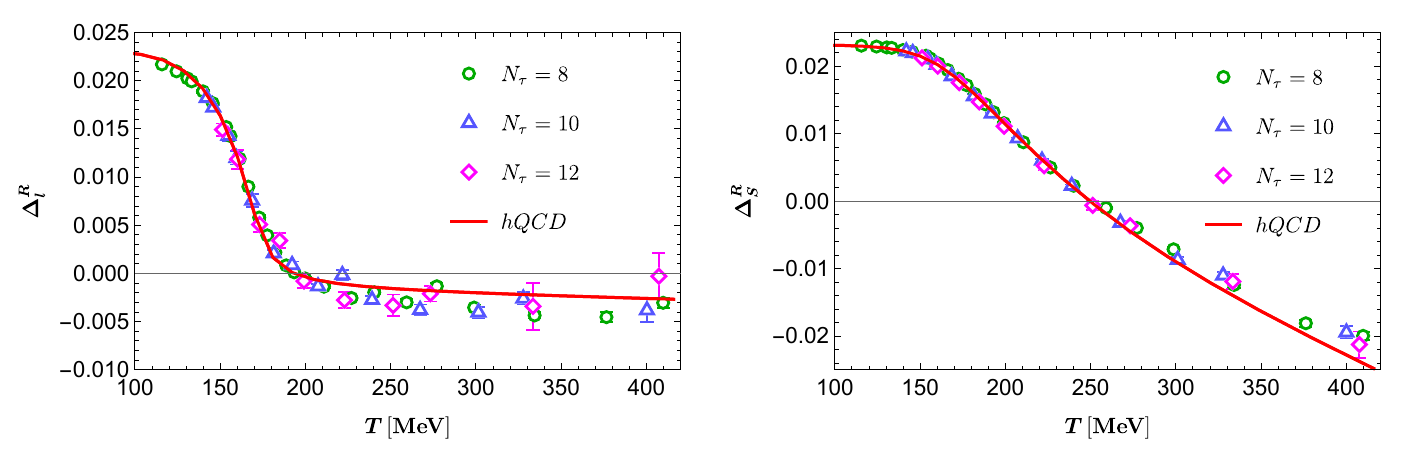}
\caption{The renormalized chiral condensates for the light quarks (left) and the $s$ quark (right). Data with error bars are from lattice QCD~\cite{HotQCD:2014kol}, while the solid curves are from our hQCD model. $N_\tau$ is the number of lattice sites along the imaginary time direction in lattice QCD.}
\label{fig:Delta}
\end{center}
\end{figure*}

The lattice simulation~\cite{Bazavov:2011nk} has applied multiplicative and additive renormalizations to define the renormalized chiral condensates $\Delta_q^R$. 
\begin{equation}\begin{aligned}\label{delta}
\Delta_q^R &=\hat{d}+2\,m_s r_1^4\Big[\langle\bar{\psi}\psi\rangle_{q,T}-\langle\bar{\psi}\psi\rangle_{q,0}\Big]\,,\text{ }\text{ }  q=l, s\,,
\end{aligned}\end{equation}
where $\hat{d} = 0.0232244$ and $r_1 = 0.3106\;\text{fm}$~\cite{HotQCD:2014kol,Bazavov:2011nk}.  The chiral condensates at zero temperature $\langle\bar{\psi}\psi\rangle_{q,0}$ can be determined from  $\langle\bar{\psi}\psi\rangle_{q,T}$ at sufficiently low temperatures. An alternative way is to consider them free parameters, determined by fitting lattice data. We find that both yield the same $\langle\bar{\psi}\psi\rangle_{q,0}$.
Our holographic results are presented in Fig.~\ref{fig:Delta} with $a_0=30, a_1=3$, and $a_l=0.595$ (for light quarks) and $a_s=1.23$ (for $s$ quark). In particular, note that both condensates change as the temperature increases. Remarkably, the holographic condensates are in quantitative agreement with those given in the lattice QCD simulation~\cite{HotQCD:2014kol}. It gives strong support for our holographic setup.

\begin{figure*}[hbt!]
\begin{center}
\includegraphics[width=1\textwidth]{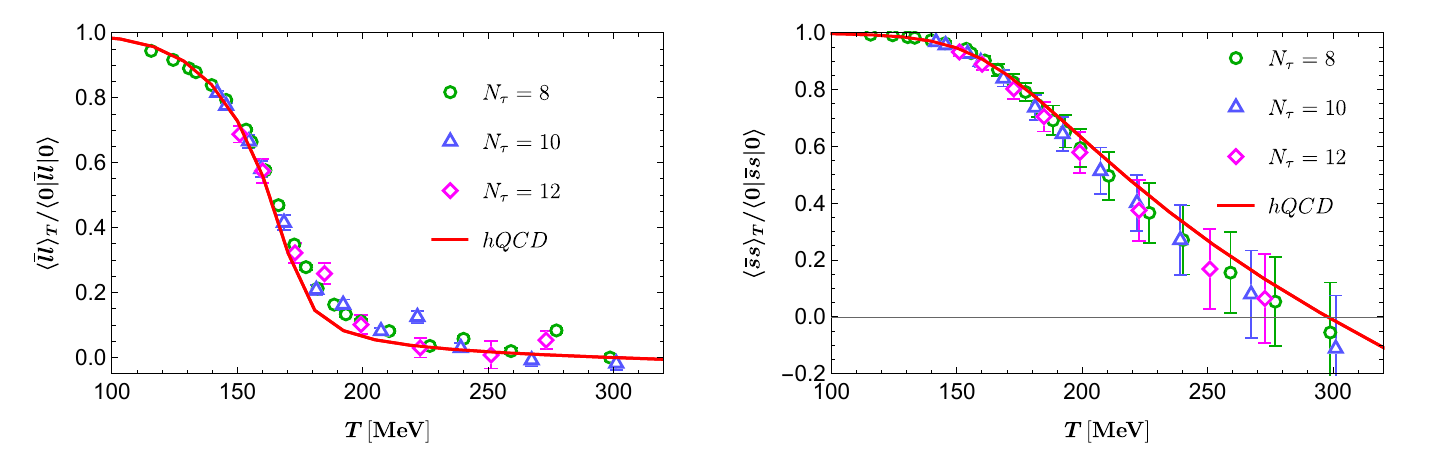}
\caption{The quark condensates for the light quarks (left) and the $s$ quark (right). Data with error bars are from lattice QCD~\cite{HotQCD:2014kol} with temporal extent $N_\tau=8, 10$, and $12$, while the solid curves from our hQCD model.}
\label{fig:qbarqT}
\end{center}
\end{figure*}

Moreover, let's consider another interesting and non-trivial check by considering the QCD trace anomaly. To do that, we should first convert the chiral condensates $\Delta_q^R$ to the quark condensates $\langle\bar{q}q\rangle_T$, which can be done by~\cite{Gubler:2018ctz}
\begin{equation}\begin{aligned}\label{vpossi14}
\frac{\langle\bar{l}l\rangle_T}{\langle0|\bar{l}l|0\rangle}&=1-\frac{\hat{d}-\Delta_l^R(T)}{\hat{d}-\Delta_l^R(\infty)}\,,\\
\frac{\langle\bar{s}s\rangle_T}{\langle0|\bar{s}s|0\rangle}&=1+\frac{\hat{d}-\Delta_s^R(T)}{2 m_s r_1^4\langle0|\bar{s}s|0\rangle}\,.
\end{aligned}
\end{equation}
For QCD with $N_f=2+1$ flavors, the vacuum quark condensates at $T=0$ take $\langle\bar{l}l\rangle_0\equiv\langle0|\bar{l}l|0\rangle=-[272(5)$MeV$]^3$ for light $u, d$ quarks and $\langle\bar{s}s\rangle_0\equiv\langle0|\bar{s}s|0\rangle=-[296(11)$MeV$]^3$ for $s$ quark~\cite{Gubler:2018ctz}. We stress that the quark condensates $\langle\bar{q}q\rangle_T$ are different from the chiral condensates $\langle\bar{\psi}\psi\rangle_{q,T}$. The quark condensates are used to evaluate the QCD trace anomaly. After choosing $\Delta_l^R(\infty)\approx\Delta_l^R(300\text{MeV})=-0.002$ by following lattice QCD~\cite{Gubler:2018ctz}, we obtain the temperature dependence of quark condensates as shown in Fig.~\ref{fig:qbarqT}. Once again, our holographic results quantitatively agree with the lattice QCD data~\cite{HotQCD:2014kol}.

The relation between the QCD trace anomaly 
$\theta(T)\equiv\epsilon(T)-3p(T)$ and the quark and gluon condensates are given by~\cite{Shifman:1978zn,Cohen:1991nk}
\begin{equation}
\begin{aligned}\label{Trelation}
\delta \langle\frac{\beta(g)}{2g}G^2\rangle_T&=\theta(T)-\hat{m}_u\delta\langle\bar{u}u\rangle_T-\hat{m}_d\delta\langle\bar{d}d\rangle_T\\
&-\hat{m}_s\delta\langle\bar{s}s\rangle_T\,,
\end{aligned}\end{equation}
where $\delta f(T)\equiv f(T)-f(0)$ denotes the vacuum subtracted value of the quantity $f$. The $\beta$-function in ~\eqref{Trelation} up to four-loop order reads~\cite{Proceedings:2019pra}
\begin{equation}\begin{aligned}\label{beta}
\frac{\beta(g)}{2g}&=-(2\pi\beta_0\alpha_s+8\pi^2\beta_1\alpha_s^2+32\pi^3\beta_2\alpha_s^3+128\pi^4\beta_3\alpha_s^4)\,,\\
\beta_0&=\frac{1}{(4\pi)^2}(11-\frac{2}{3}N_f),\quad\beta_1=\frac{1}{(4\pi)^4}(102-\frac{38}{3}N_f)\,,\\
\beta_2&=\frac{1}{(4\pi)^6}(\frac{2857}{2}-\frac{5033}{18}N_f+\frac{325}{54}N_f^2)\,,\\
\beta_3&=\frac{1}{(4\pi)^8}\Big[\frac{149753}{6}+3564\zeta_3-(\frac{1078361}{162}+\frac{6508}{27})N_f\\
&+(\frac{50065}{162}+\frac{6472}{81}\zeta_3)N_f^2+\frac{1093}{729}N_f^3\Big]\,,
\end{aligned}
\end{equation}
where the four-loop QCD running coupling $\alpha_s$ in the function of the energy scale $\mu$ is given by~\cite{Alekseev:2002zn}
\begin{equation}
\begin{aligned}\label{alphas}
\alpha_s(\mu)&=\frac{1}{4\pi\beta_0L_\mu}\Big[1-\frac{\beta_1}{\beta_0^2L_\mu}\ln(L_\mu)+\frac{\beta_1^2}{\beta_0^4L^2}\big(\ln^2(L_\mu)\\
&-\ln(L_\mu)-1+\frac{\beta_0\beta_2}{\beta_1^2}\big)-\frac{\beta_1^3}{\beta_0^6L_\mu^3}\big(\ln^3(L_\mu)-\frac{5}{2}\ln^2(L_\mu)\\
&-(2-\frac{3\beta_0\beta_2}{\beta_1^2})\ln(L_\mu)+\frac{1}{2}-\frac{\beta_0^2\beta_3}{2\beta_1^3}\big)\Big]\,,\\
L_\mu&=\ln(\mu^2/\Lambda_{QCD}^2)\,.
\end{aligned}
\end{equation}
The corresponding four-loop running quark masses read~\cite{Vermaseren:1997fq}
\begin{equation}
\begin{aligned}\label{mqmu}
\hat{m}_q(\mu)&=\bar{m}_q(\frac{\alpha_s(\mu)}{\pi})^{\gamma_0/(4\pi^2\beta_0)}\Big[1+A_1(\frac{\alpha_s(\mu)}{\pi})+\frac{(A_1^2+A_2)}{2}\\
&(\frac{\alpha_s(\mu)}{\pi})^2+\frac{1}{3}(\frac{1}{2}A_1^3+\frac{3}{2}A_1A_2+A_3)(\frac{\alpha_s(\mu)}{\pi})^3\Big]\,,\\
A_1&=-\frac{\beta_1\gamma_0}{\beta_0^2}+\frac{\gamma_1}{(2\pi)^2\beta_0},\quad\\ A_2&=(2\pi)^2\frac{\gamma_0}{\beta_0^2}(\frac{\beta_1^2}{\beta_0}-\beta_2)-\frac{\beta_1\gamma_1}{\beta_0^2}+\frac{\gamma_2}{(2\pi)^2\beta_0}\,,\\
A_3&=(2\pi)^4\Big[\frac{\beta_1\beta_2}{\beta_0}-\frac{\beta_1}{\beta_0}(\frac{\beta_1^2}{\beta_0}-\beta_2)-\beta_3\Big]\frac{\gamma_0}{\beta_0^2}\\
&+(2\pi)^2\frac{\gamma_1}{\beta_0^2}(\frac{\beta_1^2}{\beta_0}-\beta_2)-\frac{\beta_1\gamma_2}{\beta_0^2}+\frac{\gamma_3}{(2\pi)^2\beta_0}\,,\\
\gamma_0&=1,\quad \gamma_1=\frac{1}{16}(\frac{202}{3}-\frac{20}{9}N_f),\\
\quad\gamma_2&=\frac{1}{64}\Big[1249-(\frac{2216}{27}+\frac{160}{3}\zeta_3)N_f-\frac{140}{81}N_f^2\Big]\,,\\
\gamma_3&=\frac{1}{256}\Big[\frac{4603055}{162}+\frac{135680}{27}\zeta_3-8800\zeta_5\\
&+(-\frac{91723}{27}-\frac{34192}{9}\zeta_3+880\zeta_4+\frac{18400}{9}\zeta_5)N_f\\
&+(\frac{5242}{243}+\frac{800}{9}\zeta_3-\frac{160}{3}\zeta_4)N_f^2+(-\frac{332}{243}+\frac{64}{27}\zeta_3)N_f^3\Big]\,.
\end{aligned}
\end{equation}
Here $\zeta_3=1.2020569,\,\zeta_4=1.0823232,\,\zeta_5=1.0369278$ (Riemann zeta function of arguments 3, 4 and 5), and $\bar{m}_q$ is the renormalized parameter with $\bar{m}_{l}=23$MeV and $\bar{m}_s=460$MeV. In our present case the number of flavors $N_f=3$. 

\begin{figure*}[hbt!]
\begin{center}\includegraphics[width=0.7\textwidth]{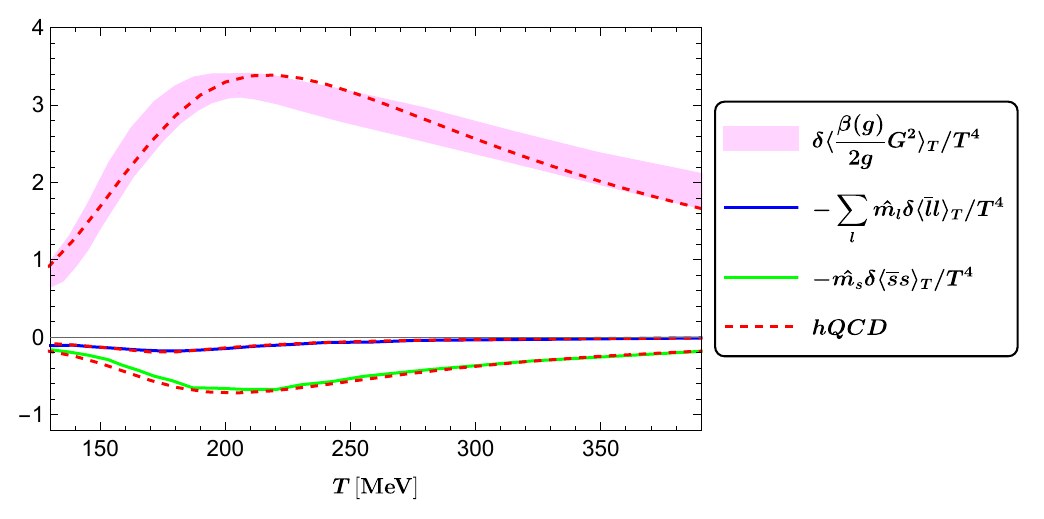}
\caption{Quark and gluon contributions to
the QCD trace anomaly. For the lattice QCD data~\cite{HotQCD:2014kol}, the pink band denotes the gluon condensate (the right-hand side of~\eqref{Trelation}), and the solid blue and red curves are, respectively, the contribution from the light $(l=u, d)$ quarks and the $s$ quark to the same equation. The corresponding results from our holographic scenario are shown as red dashed curves.}
\label{fig:Relation}
\end{center}
\end{figure*}

We have already known the temperature dependence of the quark condensates $\langle\bar{q}q\rangle_T$ using the probe scenario (see Fig.~\ref{fig:qbarqT}) and the trace anomaly $\theta$ from the EOS (see Fig.~\ref{fig:lattice_app}). Thus, once knowing the relation between the renormalization scale $\mu$ and the temperature $T$, we can obtain the gluon condensate as a function of $T$ from~\eqref{Trelation} using our holographic data. Considering that the temperature plays a natural role in a system's energy scale, it is reasonable to identify $T$ as the energy scale $\mu$ of~\eqref{mqmu}. More precisely, in our holographic setup, we choose $\mu/\Lambda_{QCD}=T/\text{MeV}$ in~\eqref{alphas}. In Fig.~\ref{fig:Relation}, the gluon condensate $\delta \langle\frac{\beta(g)}{2g}G^2\rangle_T$ from our holographic setup is denoted by the red dashed curve within the pink band given by the lattice simulation~\cite{HotQCD:2014kol}. Moreover, with the four-loop renormalization effect of quark masses~\eqref{mqmu}, the individual contributions from the light and heavy quarks to the QCD trace anomaly quantitatively agree with the lattice data, as shown in Fig.~\ref{fig:Relation}. Remarkably, such a simple holographic setup is able to capture the realist QCD thermodynamics and the condensates.\\

\textbf{Scenario with first order cosmological QCD phase transition}--The CEP predicted by our hQCD model is located at $T_C=105 \text{MeV}, \mu_C=555 \text{MeV}$. Therefore, it suggests that the strong first-order phase transition in the cosmic QCD epoch could happen for sufficiently large baryon chemical potential. However, measurements of primordial element abundances and CMB have imposed a strong constraint on the baryon asymmetry $\eta_B^{ob}\approx 10^{-10}$~\cite{Planck:2018vyg}. The baryon asymmetry is conserved except during baryogenesis and a first-order phase transition. Therefore, to apply to the cosmological QCD phase transition, the baryon asymmetry after the first-order transition in our model should be consistent with $\eta_B^{ob}$.

\begin{figure}[hbt!]
	\centering
	\includegraphics[width=0.47\textwidth]{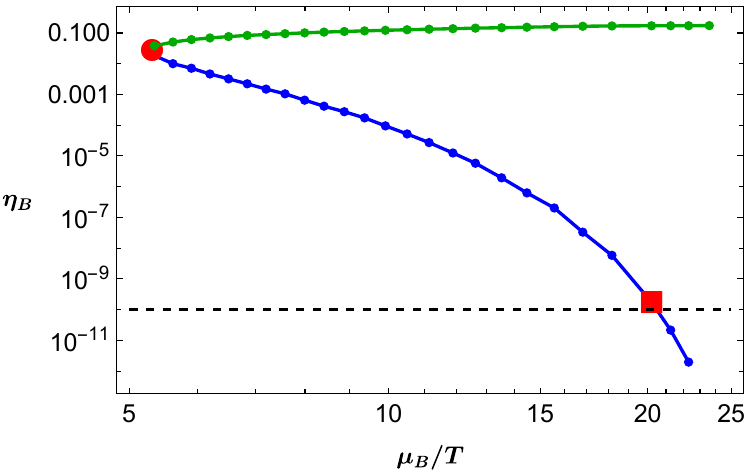}	
	\caption{The baryon asymmetry at the QCD first-order phase transition line in function of $\mu_B/T$. The blue curve is for $\eta_B$ in the false vacuum, while the red one is for the case in the true vacuum. The red dot denotes the CEP and the square is for the phase transition at $\mu_B=1000\text{MeV}$. The dashed line is the constraint on $\eta_B\approx 10^{-10}$ from observation.}
\label{fig:etaB}
\end{figure}

The value of baryon asymmetry $\eta_B$ at the QCD first-order phase transition line from our model is presented in Fig.~\ref{fig:etaB}.
While the value of $\eta_B$ in the false vacuum (green curve) slightly increases as $\mu_B/T$ increases, the one in the true vacuum (blue curve) decreases quickly. We point out that the constraint on $\eta_B$ by primordial element abundances and CMB only applies to later stages after a first-order phase transition. Thus, we need to compare our $\eta_B$ in the true vacuum at the first order transition to $\eta_B^{ob}$. Near the CEP, our model gives $\eta_B\sim 10^{-2}$ which is eight orders of magnitude larger than $\eta_B^{ob}$. Nevertheless, around $\mu_B=1000\text{MeV}$ (the red square in Fig.~\ref{fig:etaB}), $\eta_B\approx 10^{-10}$ reaches the same order of magnitude as the cosmological observation.

Therefore, the cosmological QCD phase transition happens near $\mu_B=1000\text{MeV}$ ($\mu_B/T\approx 20$) in the first order and agrees with the baryon asymmetry observation.
The GWs produced during the QCD phase transition at $\mu_B=1000\text{MeV}$ are shown in Fig.~\ref{fig:GW} in the main text. We find that it could be detected via pulsar timing in the future, such as IPTA and SKA.

\begin{figure}[hbt!]
	\centering
	\includegraphics[width=0.45\textwidth]{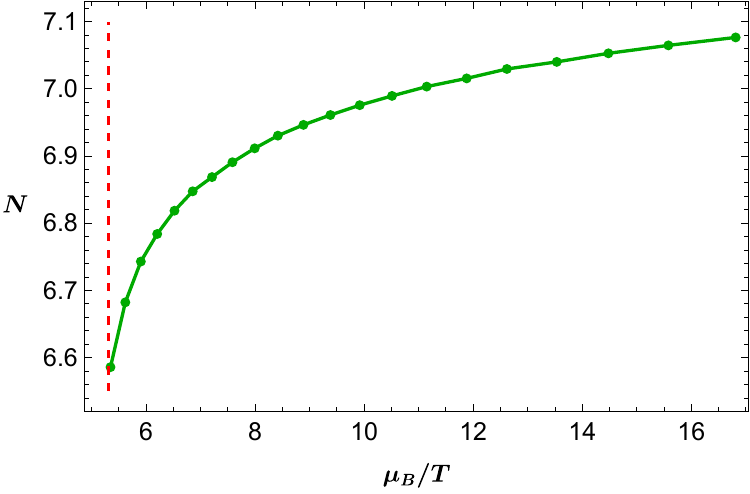}	
	\caption{The e-fold number $N$ estimated from~\eqref{eqN} as a function of $\mu_B/T$ in the little inflation scenario for our model. The vertical dashed line denotes the location of the CEP.}
\label{fig:N}
\end{figure}

Moreover, for $\mu_B<1000\text{MeV}$, there is a simple scenario with the cosmological QCD phase transition being first order without violating the constraint of a small baryon asymmetry in the later evolution of the universe. In the so-called  ``little inflation"~\cite{Boeckel:2009ej,Boeckel:2011yj} (see also~\cite{Linde:1985gh,kaempfer:1986,Borghini:2000} for earlier work), the universe begins with a large baryon asymmetry and thus crosses the first-order phase transition line. Meanwhile, the large baryon asymmetry is diluted by a short period of inflation to today's observed value. To get an order of magnitude estimate, we approximate $n_{Bi}$ ($n_{Bf}$) by $n_{B+}$ ($n_B^{ob}$) at given $\mu_B$ with $i$ and $f$ referring to the values before and after the inflation, respectively. We immediately obtain that
\begin{equation}\label{eqN}
 \frac{a_f}{a_i}=\left(\frac{\eta_{Bi}}{\eta_{Bf}}\right)^{1/3}=\left(\frac{\eta_{B+}}{\eta_B^{ob}}\right)^{1/3}\,,
\end{equation}
where $a$ is the scale factor, we have considered that the baryon number in a comoving volume is conserved. In Fig.~\ref{fig:N}, we show the corresponding e-fold number $N\equiv\ln(a_f/a_i)$ that is sufficient for the dilution of the baryon number in the little inflation scenario. We point out that this leads to an overestimation of $N$, since $n_B$ has a sudden reduction from the false to true vacuums at the first-order transition.
The e-fold number decreases by increasing $\mu_B$. Only $N\sim 7$ e-folds are sufficient.

Nevertheless, the early state of the universe has a relative large baryon asymmetry $\eta_B\sim \mathcal{O}(10^{-1})$ (see the green curve of Fig.~\ref{fig:etaB}). A natural mechanism for generating such a high initial baryon asymmetry is the well-established Affleck-Dine baryogenesis~\cite{Affleck:1984fy}. The upper limit for the Affleck-Dine baryogenesis was argued to be $\eta_B\sim \mathcal{O}(1)$~\cite{Linde:1985gh}. The case in our model is well below this upper limit.
Another scenario that allows a first-order QCD transition in agreement with observation is considering a large lepton asymmetry~\cite{Schwarz:2009ii}. A concrete realization for this scenario was recently given in~\cite{Gao:2021nwz}.

\newpage
\appendix

\end{document}